\documentstyle[aps,amssymb,amsfonts,times,mathptm,12pt]{revtex}
\begin{document}
\small
\normalsize
\protect\newtheorem{principle}{Principle} 
\protect\newtheorem{theo}[principle]{Theorem}
\protect\newtheorem{prop}[principle]{Proposition}
\protect\newtheorem{lem}[principle]{Lemma}
\protect\newtheorem{co}[principle]{Corollary}
\protect\newtheorem{de}[principle]{Definition}
\newtheorem{ex}[principle]{Example}
\newtheorem{rema}[principle]{Remark}
\newtheorem{rem}[principle]{Remark}
\newcounter{saveeqn}
\newcommand{\alpheqn}{\setcounter{saveeqn}{\value{equation}}%
\setcounter{equation}{0}%
\renewcommand{\theequation}{\mbox{\arabic{saveeqn}-\alph{equation}}}}
\newcommand{\reseteqn}{\setcounter{equation}{\value{saveeqn}}%
\renewcommand{\theequation}{\arabic{equation}}}
\renewcommand{\baselinestretch}{1}
\small
\normalsize
\title{A note on "A Matrix Realignment Method for Recognizing Entanglement,"
quant-ph/0205017 v1}
\author{Oliver Rudolph \thanks{email: rudolph@fisicavolta.unipv.it}}
\address{Quantum Optics \& Information Group, Istituto Nazionale per la Fisica
della Materia \& Dipartimento \\ di Fisica "A.~Volta", Universita di
Pavia, via Bassi 6, I-27100 Pavia, Italy}
\maketitle
\begin{abstract}
\noindent
\end{abstract}
In \cite{Chen02} Chen, Wu and Yang formulated a necessary
separability criterion based on a realignment method for matrices.
This note is to point out that this criterion is identical to the
necessary cross norm criterion put forward in
\cite{Rudolph02}. The cross norm criterion in \cite{Rudolph02} is based on a
canonical isomorphism between Hilbert-Schmidt operators
acting on ${\mathbb{C}}^n \otimes
{\mathbb{C}}^n$ and Hilbert-Schmidt operators mapping
${\mathbb{C}}^{n^2}$ into itself. The matrix realignment method
proposed in \cite{Chen02} is this canonical
isomorphism. Therefore the criteria formulated in \cite{Chen02} and
\cite{Rudolph02}
and subsequent results, including examples,
are identical. Moreover, from the
formulation of the criterion given in \cite{Rudolph02}
the statement of Proposition 1 in
\cite{Chen02} follows as an immediate corollary and it is also
immediately obvious that the dual criterion formulated in Theorem
2 in \cite{Chen02} is equivalent to the original criterion.

\end{document}